\documentclass[conference]{IEEEtran}
\IEEEoverridecommandlockouts
\usepackage{cite}
\usepackage{amsmath,amssymb,amsfonts,amsthm}
\usepackage{hyperref}
\usepackage{graphicx}
\usepackage{textcomp}
\usepackage{xcolor}
\usepackage{kotex}
\usepackage[caption=false]{subfig}

\usepackage[ruled,vlined,linesnumbered]{algorithm2e}
\usepackage{algpseudocode}

\newtheorem{lemma}{Lemma}

\newtheorem{proposition}{Proposition}

\def\BibTeX{{\rm B\kern-.05em{\sc i\kern-.025em b}\kern-.08em
    T\kern-.1667em\lower.7ex\hbox{E}\kern-.125emX}}

\begin{document}

\title{Vehicle Speed-Aware Jammer-Resilient Receiver for Multi-User MIMO V2X  Systems\\
\thanks{This work was supported by Institute of Information \& communications Technology Planning \& Evaluation (IITP) grant funded by the Korea government (MSIT) (No. RS-2024-00442085, No. RS-2024-00398157).}
}

\author{\IEEEauthorblockN{Hanyoung Park, Yongjae Jang, and Ji-Woong Choi}
\IEEEauthorblockA{Department of Electrical Engineering and Computer Science, DGIST, Daegu, South Korea \\
e-mail: \{prkhnyng, yj46.jang, jwchoi\}@dgist.ac.kr }
}

\maketitle

\begin{abstract}

Vehicle-to-everything (V2X) communication plays a crucial role in enabling connected and autonomous driving by supporting the reliable exchange of safety-critical information among vehicles and infrastructure. However, due to the open nature of wireless channels, V2X systems are vulnerable to various physical-layer attacks, among which jamming is one of the most intuitive and severe threats. In this paper, we propose a vehicle speed-aware jammer-resilient reception framework for multi-user multiple-input multiple-output (MIMO) V2X systems. The proposed method exploits the fundamental difference in Doppler characteristics between stationary jammers and moving vehicles. By transforming the received signal into the Doppler domain, the receiver identifies low-Doppler components associated with static interference and suppresses them through Doppler-domain filtering. Notably, the proposed approach does not require prior knowledge of the jammer channel or its spatial direction, making it suitable for practical V2X environments. Simulation results demonstrate that the proposed framework effectively mitigates strong jamming signals and significantly improves the achievable sum-rate compared with conventional receivers.
\end{abstract}


\section{Introduction}
With the rapid evolution of wireless communication technologies, communication networks are increasingly being applied to a wide range of services. 
In particular, in intelligent transportation systems (ITS), vehicle-to-everything (V2X) technology has emerged as a key enabler for connected and automated driving environments~\cite{ref:its}. 
V2X communication not only supports the exchange of critical driving-related information, such as vehicle location, velocity, and surrounding environmental data, but also enables the sharing of sensor data to facilitate various autonomous driving services~\cite{ref:3gpp36885, ref:channelcharting, ref:ivn}. 
Since such connectivity is closely related to road safety, ensuring real-time operation and seamless communication is of paramount importance~\cite{ref:3gpp22186, ref:latency, ref:energyconsumption}.

However, because V2X systems rely on wireless communication, they are inherently vulnerable to various security threats~\cite{ref:securitysurvey}. 
To address these challenges, considerable research has focused on enhancing the security of V2X networks~\cite{ref:secchallengesurvey}. 
Nevertheless, many existing studies have primarily focused on higher-layer security mechanisms~\cite{ref:cybersecurity, ref:ids}. 
These approaches become ineffective when the communication link itself is disrupted at the physical layer. Therefore, addressing security and reliability issues at the physical layer is essential for ensuring robust and trustworthy V2X communications.

\subsection{Prior Works}
Various studies have investigated physical layer security (PLS) techniques for V2X.
The work in~\cite{ref:key} proposed a key generation scheme that exploits the characteristics of vehicle-to-vehicle (V2V) channels. 
In~\cite{ref:analysis}, secrecy metrics for V2X PLS were analyzed. 
In addition, secure transmission techniques based on artificial noise~\cite{ref:an} and cooperative transmission~\cite{ref:cooperativejamm} have also been proposed. 
However, these approaches do not explicitly address jamming attacks, which are among the most intuitive and severe threats at the physical layer.
Several studies have considered the detection or realization of jamming attacks in V2X systems. For example, the work in~\cite{ref:jamm1} proposed a jamming detection method based on a Gaussian mixture model, while~\cite{ref:attack} demonstrated an attack strategy exploiting a reconfigurable intelligent surface. Nevertheless, these studies primarily focus on detecting or generating attacks and do not provide effective mitigation strategies against jamming.

\subsection{Contributions}
In this paper, we propose a jammer-resilient reception framework for multiple-input multiple-output (MIMO) V2X systems against jamming attacks, which are among the most intuitive yet critical threats at the physical layer. 
We first consider an uplink V2X scenario where multiple vehicles transmit to a base station in the presence of a strong jammer located near the receiver. 
While the jammer is typically positioned to deliver a strong and static interference signal, vehicles move with nonzero velocities and thus induce Doppler shifts in the received signals. 
Exploiting this fundamental difference, we devise a Doppler-aware jammer mitigation technique that suppresses strong static interference components in the received signal. 
Specifically, the proposed receiver identifies the zero-Doppler subspace associated with stationary interference and performs projection-based suppression to remove the jammer while preserving the desired signals from moving vehicles. 
Notably, the proposed method does not require prior knowledge of the jammer channel or its spatial direction, making it suitable for practical V2X environments where such information is unavailable. 
Simulation results demonstrate that the proposed approach effectively mitigates strong jamming signals and significantly improves the received SINR and achievable sum rate compared to conventional receivers.

\subsubsection*{Notation}
Boldface uppercase and lowercase letters denote matrices and vectors, respectively. 
$(\cdot)^H$ denotes Hermitian transpose, $(\cdot)^T$ means transpose, and $\mathbf{I}_N$ is $N\times N$ identity matrix. 
$\|\cdot\|$ presents $\ell_2$ norm. $\mathrm{diag}(a_1,\cdots,a_N)$ denotes a diagonal matrix with $\{a_n\}$ in its $n$-th diagonal entry. Uppercase calligraphic letters denote sets.
For a subspace $\mathcal{V}$, $\mathcal{V}^{\perp}$ denotes its orthogonal complement, i.e., the set of vectors orthogonal to all vectors in $\mathcal{V}$.

\section{System Model}\label{sec:systemmodel}
In our model, we consider a single-cell frequency-flat massive multi-user (MU)-MIMO uplink system. The base station (BS) is equipped with an $M$-element uniform linear array (ULA) to serve $K$ moving vehicles, under the presence of a single-antenna jammer.
We assume that the jammer is stationary and the vehicles are moving along the road, since a jammer located near the receiver typically generates a strong but static interference signal in practical V2X environments.
The $n$-th sample of the received signal can be modeled as
\begin{equation}
    \mathbf{y}[n] = \sum_{k=1}^K \mathbf{h}_k s_k[n] + \mathbf{h}_j s_j[n] + \mathbf{w}[n],
\end{equation}
where $\mathbf{h}_k\in\mathbb{C}^M$ is the uplink channel vector of vehicle $k$, $s_k$ is the baseband waveform of vehicle $k$, $\mathbf{h}_j\in\mathbb{C}^M$ is the jammer channel vector, $s_j$ is the jammer waveform, and $\mathbf{w}\sim\mathcal{CN}(0, \sigma_w^2 \mathbf{I}_M)$ is the noise vector.
Here, we assume that the channel remains constant over the duration of $N$ transmitted signal samples.
The channel vector of an arbitrary user or jammer is given by~\cite{ref:davidtse}
\begin{equation}
    \mathbf{h} = \sum_{\ell=1}^{L} g_{\ell} \mathbf{a}(\phi_\ell), 
\end{equation}
where $L$ is the number of dominant propagation paths including a potential line-of-sight (LoS) path, $g_\ell$ is the complex channel gain of $\ell$-th path, $\phi_\ell$ is the spatial frequency of $\ell$-th path which corresponds to angle of arrival (AoA), and $\mathbf{a}(\cdot)\in\mathbb{C}^M$ is the steering vector.
The steering vector for an arbitrary spatial frequency $\phi$ is given by
\begin{equation}
    \mathbf{a}(\phi) = \frac{1}{\sqrt{M}} [1, e^{-j\pi\phi}, e^{-j2\pi\phi},\cdots, e^{-j(M-1)\pi\phi}]^T.
\end{equation}

Also, the Doppler effect introduces a phase rotation to the received signal~\cite{ref:davidtse}. Accordingly, the waveform $s[n]$ is given by~\cite{ref:goldsmith}
\begin{equation}
    s[n] = s^\mathrm{orig}[n]\exp \left( j2\pi f_{D}n T_s \right),
\end{equation}
where $s^\mathrm{orig}$ is the original waveform, $f_D$ is the Doppler frequency, and $T_s$ is the time sampling interval. 
The Doppler frequency is expressed as
\begin{equation}
    f_D = \frac{v}{c} f_c \cos (\phi),
\end{equation}
where $v$ is the moving speed, $c$ is the speed of light, and $f_c$ is the carrier frequency.

\section{Proposed Method}
Since we assume the jammer is stationary in Sec.~\ref{sec:systemmodel}, the moving speed of the jammer is also assumed to be zero.
Thus, the Doppler frequency of the jammer $f_{D, j}$ is also zero, and it implies that $s_j[n]=s^\mathrm{orig}_j[n]$.
To suppress the jammer signal in the received signal, we utilize this characteristic to design a nullformer.
To this end, we transform the signal into the Doppler domain as follows, since signals with different velocities appear in different Doppler bins, while a stationary jammer concentrates its energy around the zero-Doppler region.
Let $N$ denote the number of time snapshots and define the received signal matrix as
\begin{equation}
    \mathbf{Y} = [\mathbf{y}[0],\mathbf{y}[1],\cdots,\mathbf{y}[N-1]].
\end{equation}
Applying the discrete Fourier transform (DFT) yields the Doppler-domain representation, which is derived as
\begin{equation}
    \bar{\mathbf{Y}} = \mathbf{YF}^H,
\end{equation}
where $\mathbf{F}\in\mathbb{C}^{N\times N}$ is unitary DFT matrix and each column of $\bar{\mathbf{Y}}$ corresponds to a specific Doppler bin.

Here, we assume $f_{D,j}$ is zero, and it means that its energy resides in Doppler bins around zero frequency. Therefore, we suppress the zero-Doppler subspace by applying a diagonal Doppler-domain mask, which is given by
\begin{equation}\label{eqn:masking}
    \mathbf{M} = \mathrm{diag}(m_1,\cdots,m_N),
\end{equation}
where
\begin{equation}
    m_n=\begin{cases}
        & 0,\quad |f_n|\leq f_\mathrm{cutoff}, \\
        & 1,\quad \textnormal{otherwise,}
    \end{cases}
\end{equation}
with the Doppler bin frequency $f_n$ and the Doppler cutoff bandwidth $f_\mathrm{cutoff}$. 
This mask removes the basis corresponding to the low-Doppler subspaces.
Note that the Doppler bin frequency is given by
\begin{equation}
    f_n = \frac{n}{NT_s}.
\end{equation}
Subsequently, it is applied to the Doppler-domain received signal sample as
\begin{equation}
    \bar{\mathbf{Y}}^\mathrm{filtered} = \bar{\mathbf{Y}}\mathbf{M}.
\end{equation}
However, since this operation is defined in the Doppler domain and using DFT for every case can be computationally burdensome, it can be simplified as follows because $\mathbf{M}$ is constant for a desired $f_\mathrm{cutoff}$, which is derived as
\begin{equation}
    \mathbf{Y}^\mathrm{filtered} 
    = \bar{\mathbf{Y}}^\mathrm{filtered}\mathbf{F}
    = \bar{\mathbf{Y}}\mathbf{M}\mathbf{F}
    = \mathbf{Y}\mathbf{F}^H\mathbf{M}\mathbf{F}
    = \mathbf{Y}\mathbf{P},
\end{equation}
where $\mathbf{P}=\mathbf{F}^H\mathbf{M}\mathbf{F}$ is the masking matrix in time domain.
Proposition~\ref{prop:1} shows that the masking matrix in the time domain is equivalent to the orthogonal projection onto the nullspace of the Doppler bins corresponding to low Doppler frequencies.
\begin{proposition}\label{prop:1}
    Let $\mathbf{M}$ be the masking matrix defined in ~\eqref{eqn:masking} and define
    \begin{equation}
        \mathbf{P} = \mathbf{F}^H \mathbf{MF}.
    \end{equation}
    Then $\mathbf{P}$ is an orthogonal projection matrix that removes the Doppler subspace spanned by Doppler bins satisfying $|f_n|\leq f_\mathrm{cutoff}$.
\end{proposition}
\begin{proof}
    Please refer to Appendix~\ref{app:1}.
\end{proof}

Subsequently, since the received signal still has distortions from the MIMO uplink channel, its equalization is conducted with a linear minimum mean-squared error (LMMSE) equalizer, which is a broadly considered linear equalizer for MIMO uplink systems, and its filter matrix is given by
\begin{equation}
    \mathbf{V}_\mathrm{LMMSE} = \widehat{\mathbf{H}}(\widehat{\mathbf{H}}^H\widehat{\mathbf{H}}+\sigma^2_w\mathbf{I}_K)^{-1},
\end{equation}
where $\widehat{\mathbf{H}}\in\mathbb{C}^{M\times K}$ is the estimated uplink channel matrix.
Using this filter matrix, the symbol detection is performed as
\begin{equation}
    \hat{\mathbf{s}}[n] = \mathbf{V}^H\mathbf{y}[n].
\end{equation}

\section{Performance Evaluation}

\subsection{Simulation Setup}
We consider the 32-element ULA with half-wavelength antenna spacing. The BS serves 6 vehicles and each vehicle is assumed to be equipped with a single omnidirectional antenna.
For the uplink channel, we adopt the QuaDRiGa channel model corresponding to the 3GPP 38.901 UMi scenario~\cite{ref:quadriga, ref:3gpp38901} with carrier frequency of 1.9 GHz. The per-user SNR is assumed to vary within -3 to 3 dB around the average SNR, and we assume that the jammer power in the received signal is 30 dB higher than the average user receive power.
Additionally, the jammer is assumed to be placed at a location where a LoS path to the BS is available. The AoA of the LoS path $\phi_{j,\mathrm{LoS}}$ follows the uniform distribution in the range $[-40^\circ, 40^\circ]$.
The vehicle speed relevant to the Doppler effect is defined as the absolute effective velocity, given by
\begin{equation}
    v_{\mathrm{eff}} = |v \cos(\phi)|,
\end{equation}
where $v$ denotes the vehicle velocity in the two-dimensional plane and $\phi$ represents the angle between the moving direction of the vehicle and the array broadside.
The effective speed of each vehicle is generated according to a uniform distribution around the average effective speed $\bar{v}_{\mathrm{eff}}$, i.e.,
\begin{equation}
 v_{\mathrm{eff}} \sim \mathcal{U}(\bar{v}_{\mathrm{eff}} - \Delta_v,\; \bar{v}_{\mathrm{eff}} + \Delta_v),   
\end{equation}
where $\Delta_v$ is 10 km/h.
Simulation results are obtained from $10^4$ Monte Carlo trials.
Additionally, we assume perfect channel estimation for simplicity, as prior works on channel estimation has demonstrated remarkable accuracy~\cite{ref:channelest1, ref:channelest2, ref:channelest3}. For the received signal, we assume that the waveforms of the vehicles and the jammer follow zero-mean complex Gaussian distributions. 

For the performance metric, we consider the sum-rate using the signal-to-interference-plus-noise ratio (SINR) in consideration of the jammer as interference, which is defined as
\begin{equation}
    \sum_{k=1}^K \log_2 (1+ \mathrm{SINR}_k),
\end{equation}
where the SINR of each user is defined as~\cite{ref:sinr}
\begin{equation}
\begin{aligned}
    & \mathrm{SINR}_k \\
    & = \frac{\|\mathbf{v}_k^H\mathbf{h}_k \mathbf{s}_k\mathbf{P}\|^2} {\sum_{k^\prime \neq k} \|\mathbf{v}_k^H \mathbf{h}_{k^\prime} \mathbf{s}_{k^\prime}\mathbf{P}\|^2 
    + \|\mathbf{v}_k^H \mathbf{h}_j \mathbf{s}_j\mathbf{P}\|^2 + \|\mathbf{v}_k^H\mathbf{W}\mathbf{P}\|^2},
\end{aligned}
\end{equation}
where $\mathbf{s}_k\in\mathbb{C}^{1\times N}$ and $\mathbf{s}_j\in\mathbb{C}^{1\times N}$ are the transmitted waveform snapshots of user $k$ and jammer, respectively, which are defined as
\begin{equation}
    \begin{aligned}
        \mathbf{s}_k = [s_k[0],s_k[1],\cdots,s_k[N-1]],\\
        \mathbf{s}_j = [s_j[0],s_j[1],\cdots,s_j[N-1]],\\
    \end{aligned}
\end{equation}
and $\mathbf{W}\in\mathbb{C}^{M\times N}$ is the received noise snapshots as
\begin{equation}
    \mathbf{W} = [\mathbf{w}[0],\mathbf{w}[1],\cdots,\mathbf{w}[N-1]].
\end{equation}
Additionally, to compare with the unmitigated case, the SINR is calculated as
\begin{equation}
\begin{aligned}
    &\mathrm{SINR}_k^\textnormal{unmitigated}  \\
    &= \frac{\|\mathbf{v}_k^H\mathbf{h}_k \mathbf{s}_k\|^2} {\sum_{k^\prime \neq k} \|\mathbf{v}_k^H \mathbf{h}_{k^\prime} \mathbf{s}_{k^\prime}\|^2 
    + \|\mathbf{v}_k^H \mathbf{h}_j \mathbf{s}_j\|^2 + \|\mathbf{v}_k^H\mathbf{W}\|^2}.
\end{aligned}
\end{equation}

\subsection{Simulation Results}
\begin{figure}
    \centering
    \includegraphics[width=0.9\linewidth]{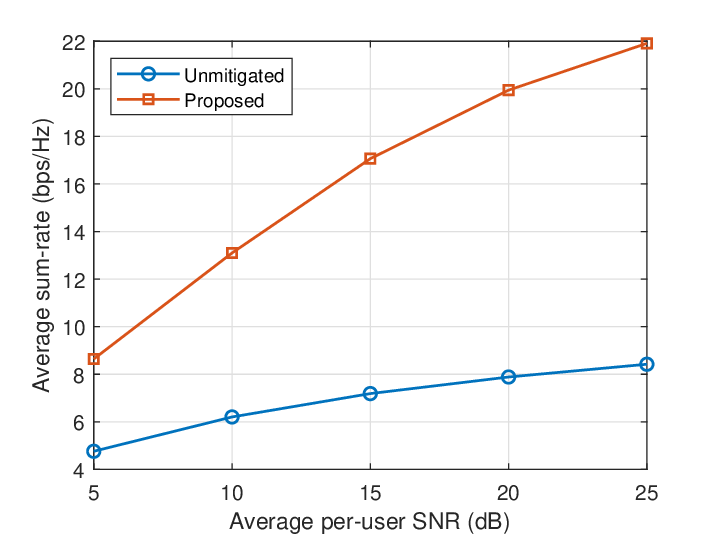}
    \caption{Sum-rate depending on the average received SNRs.}
    \label{fig:snrsumrate}
\end{figure}

Fig.~\ref{fig:snrsumrate} shows the post-filter sum-rate depending on the average received SNR when the average vehicle speed is 30 km/h. In the unmitigated case, the sum-rate increases only marginally as the SNR increases due to the strong interference from the jammer. In contrast, the proposed method achieves a substantial sum-rate improvement as the SNR increases. This behavior can be explained by the fact that lower noise power improves the separability of signal components in the Doppler domain, thereby enhancing the effectiveness of jammer suppression.

\begin{figure}
    \centering
    \includegraphics[width=0.9\linewidth]{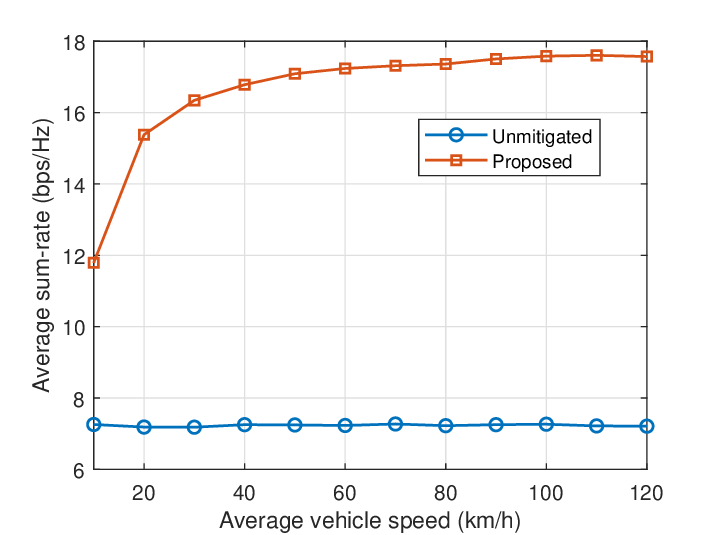}
    \caption{Sum-rate depending on the average vehicle speeds.}
    \label{fig:speedsumrate}
\end{figure}

Fig.~\ref{fig:speedsumrate} illustrates the post-filter sum-rate depending on the average vehicle speed when the average SNR is 15 dB. 
As the vehicle speed increases, the Doppler shifts of the vehicle signals become larger, pushing their spectral components farther from the Doppler bins within the cutoff frequency. Consequently, a smaller portion of the desired signal is removed by the Doppler-domain filtering, while the stationary jammer remains concentrated near the zero-Doppler region. This results in improved preservation of the desired signals and leads to higher achievable sum-rates.

\section{Conclusion}
\subsection{Summary}
In this paper, we proposed a vehicle speed-aware jammer-resilient reception framework for multi-user MIMO V2X systems. The proposed approach exploits the fundamental difference in Doppler characteristics between stationary jammers and moving vehicles. By transforming the received signal into the Doppler domain, the receiver suppresses low-Doppler components associated with static interference while preserving the desired signals from moving vehicles. Notably, the proposed method does not require prior knowledge of the jammer channel or its spatial direction, which makes it suitable for practical V2X environments. Simulation results demonstrate that the proposed framework effectively mitigates strong jamming signals and significantly improves spectral efficiency.

\subsection{Limitations and Future Work}
In this work, the jammer is assumed to be stationary, and therefore the proposed method is applicable only to scenarios where the jammer is stationary or moving at a low speed. Moreover, time-varying AoA was not considered in this study, which is more general in practical V2X systems. Accordingly, extending the proposed approach to handle dynamically changing AoA over time is our ongoing future work.

\appendices
\section{Proof of Proposition~\ref{prop:1}}\label{app:1}
Let the Doppler subspace corresponding to the Doppler bins within the cutoff region be defined as
\begin{equation}
    \mathcal{S}=\mathrm{span} \{ \mathbf{f}_n:|f_n|\leq f_\mathrm{cutoff} \},
\end{equation}    
where $\mathbf{f}_n$ is the $n$-th column of the unitary DFT matrix $\mathbf{F}$.
To prove this proposition, we first establish the following lemma.
\begin{lemma}\label{lem:1}
    Consider masking matrix defined in ~\eqref{eqn:masking} and $\mathbf{P}=\mathbf{F}^H\mathbf{MF}$. Then, $\mathbf{P}$ is an orthogonal projection matrix.
\end{lemma}
\begin{proof}
    Since $\mathbf{M}$ is a diagonal matrix whose entries are either 0 or 1, it satisfies
    \begin{equation}
        \mathbf{M}^2=\mathbf{M},
    \end{equation}
    and
    \begin{equation}
        \mathbf{M}^H = \mathbf{M}.
    \end{equation}
    Also, because the DFT matrix $\mathbf{F}$ is unitary, we have
    \begin{equation}
        \mathbf{F}\mathbf{F}^H = \mathbf{I}_N.
    \end{equation}
    Using these properties,
    \begin{equation}
    \begin{aligned}
        \mathbf{P}^2 &= \mathbf{F}^H_N\mathbf{MF}\mathbf{F}^H\mathbf{MF}\\
        &= \mathbf{F}^H \mathbf{M}^2\mathbf{F}\\
        &= \mathbf{F}^H \mathbf{MF}.\\
    \end{aligned}
    \end{equation}
    Thus,
    \begin{equation}
        \mathbf{P}^2 = \mathbf{F}^H \mathbf{MF}= \mathbf{P},
    \end{equation}
    which shows that $\mathbf{P}$ is idempotent.
    Moreover, $\mathbf{P}^H$ can be expressed as
    \begin{equation}
    \begin{aligned}
        \mathbf{P}^H & = (\mathbf{F}^H\mathbf{MF})^H \\
        &= \mathbf{F}^H\mathbf{M}^H\mathbf{F} \\
        &= \mathbf{F}^H\mathbf{MF} \\
        &= \mathbf{P}.
    \end{aligned}
    \end{equation}
    Hence, $\mathbf{P}$ is Hermitian.
    Accordingly, $\mathbf{P}$ is an orthogonal projection matrix, since $\mathbf{P}$ is idempotent and Hermitian.
\end{proof}
Let an arbitrary vector $\mathbf{z}$ be expanded in the Doppler basis as
\begin{equation}
    \mathbf{z} = \sum_{n=1}^N \alpha_n \mathbf{f}_n,
\end{equation}
where $\alpha_n$ is an arbitrary coefficient.
Applying $\mathbf{P}$ to $\mathbf{z}$ yields
\begin{equation}
    \mathbf{Pz} = \mathbf{F}^H\mathbf{MF}\mathbf{z},
\end{equation}
which corresponds to removing the Doppler coefficients within the cutoff region. Thus,
\begin{equation}
    \mathbf{Pz} = \sum_{|f_n|>f_\mathrm{cutoff}} \alpha_n\mathbf{f}_n.
\end{equation}
If $\mathbf{z}\in\mathcal{S}$, then
\begin{equation}
    \mathbf{z} = \sum_{|f_n|\leq f_\mathrm{cutoff}} \alpha_n \mathbf{f}_n,
\end{equation}
which implies 
\begin{equation}
    \mathbf{Pz}=0.
\end{equation}
Hence,
\begin{equation}
    \mathrm{Null}(\mathbf{P}) = \mathcal{S}.
\end{equation}
Since Lemma~\ref{lem:1} shows that $\mathbf{P}$ is an orthogonal projection matrix, this implies that $\mathbf{P}$ projects the received signal onto the orthogonal complement $\mathcal{S}^{\perp}$. Therefore, the operator $\mathbf{P}$ removes the signal component lying in the cutoff Doppler subspace. \qed

\bibliographystyle{IEEEtran}
\bibliography{IEEEabrv, bibliography}

\end{document}